\documentclass[prl,aps,twocolumn,showpacs]{revtex4}

\usepackage{epsfig}
\usepackage{latexsym}
\usepackage{graphicx}

\newcommand{\bi}{\begin{itemize}}
\newcommand{\ei}{\end{itemize}}
\newcommand{\be}{\begin{equation}}
\newcommand{\ee}{\end{equation}}

\newcommand{\bea}{\begin{eqnarray}}
\newcommand{\eea}{\end{eqnarray}}
\newcommand{\beastar}{\begin{eqnarray*}}
\newcommand{\eeastar}{\end{eqnarray*}}

\newcommand{\eq}[1]{(\ref{#1})}
\newcommand{\eqq}[2]{(\ref{#1},\ref{#2})}

\begin{document}

\title{Force dependent fragility in RNA hairpins}

\author{M. Manosas$^1$, D. Collin$^2$ and F. Ritort$^1$}

\affiliation{$^1$ Departament de F\'{\i}sica Fonamental, Facultat de
F\'{\i}sica, Universitat de Barcelona Diagonal 647, 08028 Barcelona, Spain\\
$^2$ Merck \& Co. Inc., Automated Biotechnology Dpt., North Wales PA
19454, USA}
\date{\today}

\begin{abstract}
We apply Kramers theory to investigate the dissociation of
multiple bonds under mechanical force and interpret experimental
results for the unfolding/refolding force distributions of an RNA
hairpin pulled at different loading rates using laser tweezers. We
identify two different kinetic regimes depending on the range of
forces explored during the unfolding and refolding process. The
present approach extends the range of validity of the two-states
approximation by providing a theoretical framework to reconstruct
free-energy landscapes and identify force-induced structural changes
in molecular transition states using single molecule pulling
experiments. The method should be applicable to RNA hairpins with
multiple kinetic barriers.
\end{abstract}

\pacs{82.35.-x,82.37.-j,87.15.-v}

\maketitle

Single molecule pulling experiments allow to exert
mechanical force on individual molecules such as nucleic acids, proteins
and macromolecular complexes \cite{Review}. By recording force-extension curves it is
possible to determine free energies and kinetic parameters of
biomolecules and search for intermediates and pathways in
biochemical reactions. Over the past years single molecule techniques have been
successfully applied to investigate the breakage of molecular bonds in
many biological systems such as proteins \cite{Mariano}, DNA molecules
\cite{bockelmann97}, RNA molecules \cite{Liphardt01}, ligand-receptor binding \cite{fritz98} and beyond, e.g. metallic gold nanowires stretched
with AFM \cite{Agrait01}. 
Under mechanical load all these structures yield at
different values of the applied force in a dynamical process that is stochastic and loading
rate dependent. The study of breakage forces under nonequilibrium conditions is known as
dynamic force spectroscopy \cite{EvansWilliams01}.
A detailed comprehension of the rupture kinetics of biomolecular
complexes has implications in our understanding of their kinetic
stability which is important in enzymatic and/or regulatory
processes. 

\begin{figure}
\begin{center}
\includegraphics[scale=0.3]{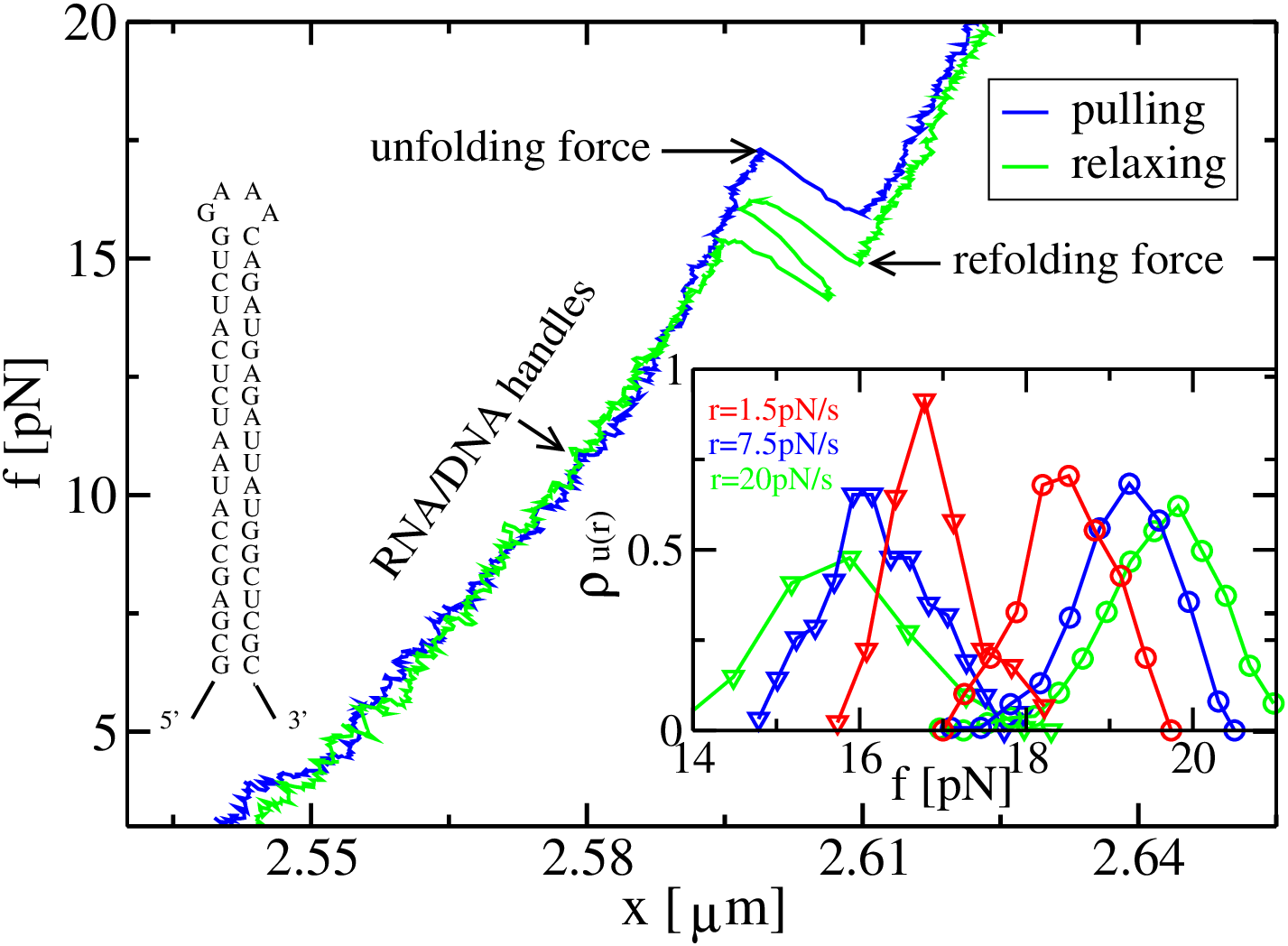}
\caption{\small{ Experimental FEC for the RNA hairpin in a ramping cycle
where the force is first increased and after decreased at a rate of $r=3
\rm pN/s$, the blue and green curves representing the pulling and
relaxing processes respectively. The arrows indicate the unfolding
and refolding forces, which correspond to the {\em first} 
unfolding and refolding events.  The first part of the FEC corresponds to the stretching 
of the RNA/DNA hybrid handles. Inset: Experimental distribution of 
the unfolding (circles) refolding (triangles) forces. Number of pulls: 
129, 385 and 703 pulls for $r$=1.5, 7.5 and 20 pN/s (red, blue and green)
respectively.}}
\label{f0}
\end{center}
\end{figure}

Here we investigate the unfolding/refolding kinetics of RNA hairpins
using laser tweezers \cite{Liphardt01,Smith03}. The RNA sequence and its native structure are shown
in Fig.\ref{f0}. To manipulate the RNA hairpin two beads are attached
to the ends of the RNA hairpin by inserting two hybrid RNA/DNA handles
\cite{foot1}.  One of the beads is immobilized on the tip of a
micropipette while the other bead is captured in the optical trap. By
moving the micropipette a force is exerted upon the ends of the RNA
hairpin and the force-extension curve (FEC) recorded.  In Fig.\ref{f0}
we show the experimental FEC corresponding to a complete cycle of a
ramping process where the force is first raised and relaxed
afterwards. In the pulling process the molecule is initially in its
native folded structure, and the force is increased at a certain rate
$r$ until the molecule unfolds. If the process is reversed, i.e. the
force is decreased at rate $-r$, the molecule folds back again
(relaxing process) \cite{Manosas05}. The unfolding/refolding 
of the molecule
can then be identified as force-extension jumps observed in the FEC.
By repeatedly pulling the molecule many times we obtain the distribution of unfolding (refolding) forces,
i.e. the force at which the {\em first} unfolding (refolding) event occurs along the
pulling (relaxing) process (Fig.\ref{f0}). The experimental distribution
of the unfolding (u) and refolding (r) forces $\rho_{\rm {u(r)}}(f)$ at
different loading rates $r$ is shown as an inset of Fig.\ref{f0}.

To model the hairpin we follow Cocco et al. \cite{cocco03} and restrict the number of
configurations of an $N$ base-pair (bp) RNA hairpin to the set of
configurations where the first $n$ bps are opened and the last $N-n$
are closed (the total number of configurations being $N+1$). The
end-to-end distance is a well defined reaction coordinate for the
unfolding/refolding reaction. We use the variable $n$ to label the
state of the hairpin; e.g the folded (F) state corresponds to $n=0$
and the unfolded (UF) state to $n=N$. The stability of each state $n$
depends on its free energy, $G_{n}(f)$, at a given applied force $f$ 
\cite{cocco03},
\be
 G_{n}(f)=G^{0}(n)+g(n,f)
\label{gn}
\ee
where $G^{0}(n)$ is the free energy of formation at
zero force and $g(n,f)$ is the force-dependent contribution to the
free energy. The latter is given by
$g(n,f)=G_{\rm{ss}}(n,f)-fx_{n}(f)$ where the functions $x_{n}(f)$ and
$G_{\rm{ss}}(n,f)$ are the end-to-end distance and the entropic
correction to the free energy of an $M$-bases long single stranded RNA
(ssRNA) at force $f$, $M$ being the number of bases released after the
opening of $n$ bps. The latter can be computed as the reversible mechanical work
needed to stretch the ends of an $M$-bases long ssRNA a distance
$x_{n}(f)$,
\be
G_{\rm{ss}}(n,f)=\int_{0}^{x_{n}(f)}F_{\rm ssRNA}(x)dx~,
\label{gr}
\ee
where $F_{\rm ssRNA}(x)$ is the FEC of the ssRNA \cite{foot2}. 
The free energy landscape of a hairpin, $G_{n}(f)$, as a function of $n$
is known to be rugged with different kinetic barriers
(or transition states) depending on the sequence and on the applied
force $f$~\cite{Nelson}. We use the Mfold prediction \cite{Mfold}
to extract the free energy of the
molecule at zero force, $G^{0}(n)$. In what follows we
take the F state as the reference state for the free energy, i.e.
$G_{n=0}(f)=G^{0}(n=0)=0$. 

The kinetics of unfolding (i.e. the transition between the
completely folded (F) and the completely unfolded (UF) states) is an
activated process with a force-dependent effective barrier, $B_{\rm
eff}(f)$, measured relative to the F state.  The rates of unfolding
and refolding, $k_{\rm{u}}(f)$ and $k_{\rm{r}}(f)$, can be computed as the
first passage rates \cite{Zwanzig} for a Brownian particle to
cross a force-dependent effective barrier $B_{\rm eff}(f)$: 
%
\begin{eqnarray}
k_{\rm{u}}(f)=k_{0}e^{-\beta B_{\rm eff}(f)};
k_{\rm{r}}(f)=k_{0}e^{-\beta (B_{\rm eff}(f)-G_{N}(f))},
\label{e8}
\end{eqnarray}
where $G_{N}(f)$ is the free energy difference between the F and UF
states at force $f$, $\beta=1/k_{B}T$ with $k_{B}$ and $T$ being
respectively the Boltzmann constant and the bath temperature, and
$k_{0}$ is an attempt frequency. An analytical expression for $B_{\rm
eff}(f)$ can be derived from Kramers theory applied to the
dissociation of consecutive bonds under mechanical force in the
stationary approximation \cite{EvansWilliams01,Zwanzig},
\begin{eqnarray}
B_{\rm eff}^{\rm KT}(f)=k_{B}T \ln \Big{[}\sum_{n=0}^{N}h(n)e^{\beta
G_{n}(f)}\Big{]},
\label{bef}
\end{eqnarray}
with $h(n)=\sum_{n'=0}^{n}e^{-\beta G_{n'}(f)}$. The variation in force
of the effective barrier gives information about its position along the
reaction coordinate:
\begin{eqnarray}
x^{\rm{F}}_{\rm eff}(f)=-\frac{\partial B_{\rm eff}(f)}{\partial f}~,~x^{\rm{UF}}_{\rm eff}(f)=\frac{\partial(B_{\rm eff}(f)-G_{N}(f))}{\partial f},
\label{xef}
\end{eqnarray}
where $x^{\rm{F}}_{\rm eff}(f)$ and $x^{\rm{UF}}_{\rm eff}(f)$ are the
distances from the effective barrier to the F and UF states
respectively. The location of the barrier along the reaction
coordinate is related to the fragility of the molecule
which determines how much the unfolding/refolding
kinetics is sensitive to the force. To characterize the fragility 
we introduce the parameter $\mu$ defined as:
\begin{eqnarray}
\mu(f)&=&\frac{x^{\rm{F}}_{\rm eff}(f)-x^{\rm{UF}}_{\rm eff}(f)}{x^{\rm{F}}_{\rm eff}(f)+x^{\rm{UF}}_{\rm eff}(f)}~;
\label{mu}
\end{eqnarray}
$\mu<0$ corresponds to a brittle structure -e.g. the case of hairpins
stabilized by tertiary contacts where the barrier is located near to
the F state-, whereas $\mu>0$ represents a flexible or compliant
structure, i.e.  molecules that can easily deform under applied force
and the barrier is close to the UF state \cite{leffler}. 

\begin{figure}
\begin{center}
\includegraphics[scale=0.3]{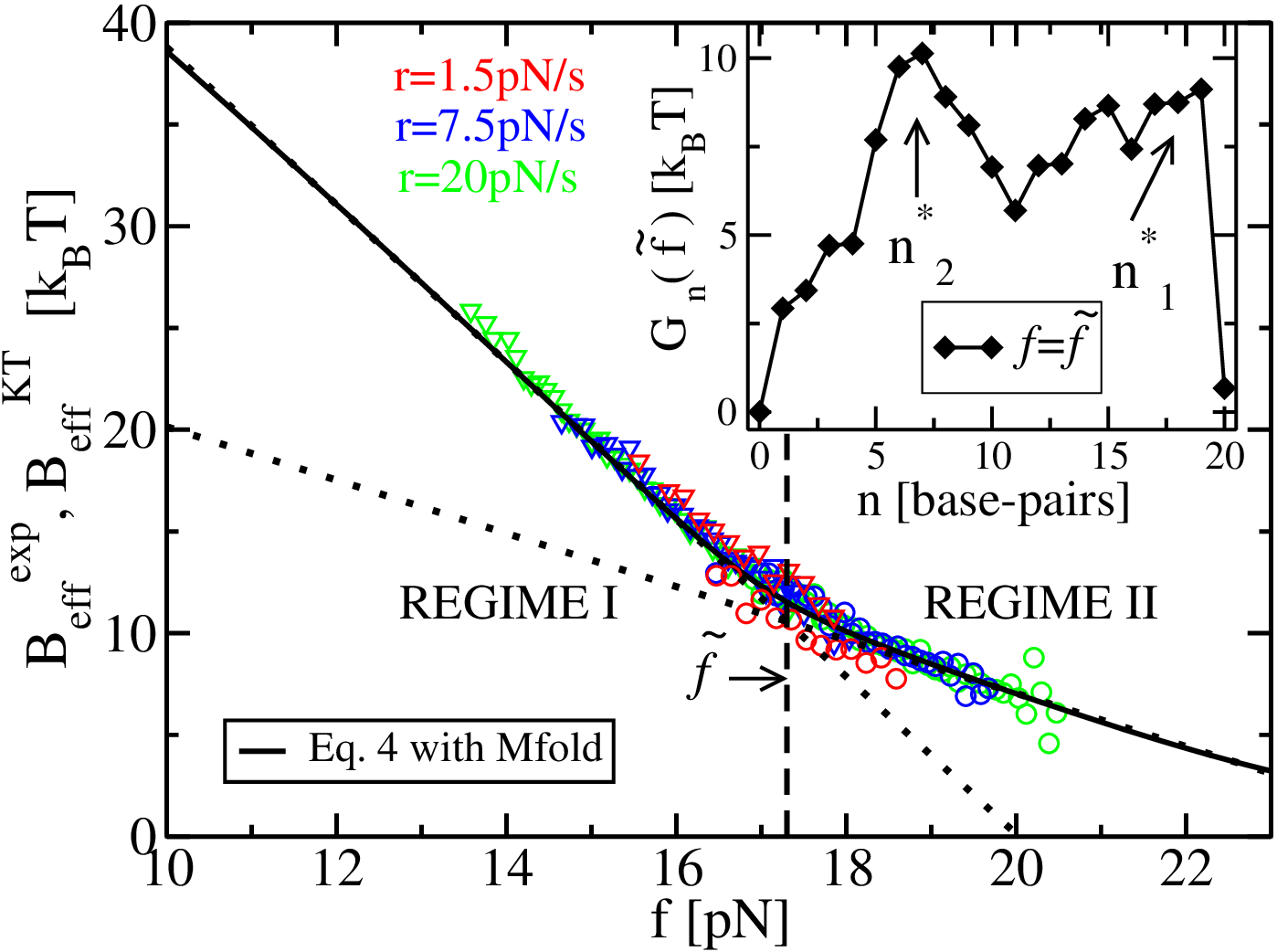}
\caption{\small{ Effective barrier $B_{\rm eff}^{\rm
exp}(f)$ \eqq{befe1}{befe2} from the pulling (circles) and relaxing
(triangles) experimental data compared with the prediction by Kramers theory, $B_{\rm
eff}^{\rm KT}$ \eq{bef} (continuous line). Inset: Free energy
landscape \eq{gn} of the RNA hairpin at the crossover force
$\tilde{f}=17.3\rm pN$.}}
\label{f3}
\end{center}
\end{figure}

The rates \eq{e8} are
related to the force distributions by the expression
$\rho_{\rm{u(r)}}(f)= \frac{k_{\rm{u(r)}}(f)}{r}\exp[${-\tiny{(+)}}
$\int_{f'}^{f}\frac{k_{\rm{u(r)}}(y)}{r}dy]$ with $f'$ being the initial
force in the pulling (relaxing) process \cite{evans}. The
unfolding (refolding) rates read as
$k_{\rm{u(r)}}(f)=\rho_{\rm{u(r)}}(f)\frac{1} {rP_{\rm u(r)}(f)}$, where
$P_{\rm u(r)}(f)$ is the probability that the molecule remains in the F
(UF) state along the pulling (relaxing) process until reaching
the force $f$, $P_{\rm
u(r)}(f)=${\small{+(-)}}$\int_{f'}^{f}\rho_{\rm{u(r)}}(y)dy$. Note that
the experimental FECs show a force jump $\delta f$ (Fig.\ref{f0})
when the molecule unfolds or refolds that corresponds to the relaxation
of the bead in the trap after the sudden increase or decrease in the RNA
extension. To compensate for this effect we shift the value of the
folding forces by an amount equal to {\small{+(-)}}$\delta f/2$.
From \eqq{gn}{e8} and the unfolding (refolding) force distributions, $\rho_{\rm{u(r)}}(f)$, 
we can extract the effective barrier as:
\begin{eqnarray}
B_{\rm eff}^{\rm exp}(f-\frac{\delta f}{2})=-\frac{1}{\beta}\ln
\Big{[}\frac{\rho_{\rm{u}}(f)r}{P_{\rm u}(f)k_{0}}\Big{]}, \label{befe1}
\end{eqnarray}
\begin{eqnarray}
B_{\rm eff}^{\rm exp}(f+\frac{\delta f}{2})=G^{0}(N)+g(N,f)-\frac{1}{\beta}\ln
\Big{[}\frac{\rho_{\rm{r}}(f)r}{P_{\rm r}(f)k_{0}}\Big{]}.
\label{befe2}
\end{eqnarray}
Using polymer theory \cite{foot2} we can estimate $g(n,f)$ so the
expressions \eqq{befe1}{befe2} have only two unknown parameters,
$G^{0}(N)$ and $k_0$.  We determine $G^{0}(N)$ by collapsing into a
single curve the effective barrier estimates \eq{befe1} and \eq{befe2}
corresponding to the pulling and relaxing processes at different
loading rates. From our data we get $G^{0}(N)=64.5k_{B}T$ in very good
agreement with the Mfold prediction, $G^{0}_{\rm Mfold}(N)=38{\rm
Kcal/mol}=63k_{B}T$ \cite{Mfold}. In Fig.\ref{f3} we show the force
dependent effective barrier obtained using this method. We then determine
the value of the attempt frequency $k_{0}$ by fitting
\eqq{befe1}{befe2} to the prediction by Kramers theory \eq{bef}. We obtain
$k_{0}=10^{5}\rm{s}^{-1}$ which is of the order of magnitude of the
values reported for other hairpins \cite{thirum}. The
agreement found between the predicted effective barrier \eq{bef} and
the results from the experiments \eqq{befe1}{befe2} validates our
description hence providing a way to estimate the attempt frequency
$k_{0}$~\cite{foot5}.  The representation of the effective barrier as
a function of the applied force reflects two distinct regimes
(Fig.\ref{f3}) characterized by different slopes of $B_{\rm eff}(f)$.
These correspond to different locations of the effective barrier
\eq{xef} and different values of the fragility \eq{mu}.  We define a
crossover force $\tilde{f}$ as the value at which the extrapolated
straight lines corresponding to regimes I and II intersect each other
(Fig.\ref{f3}).

\begin{figure}
\begin{center}
\includegraphics[scale=0.3]{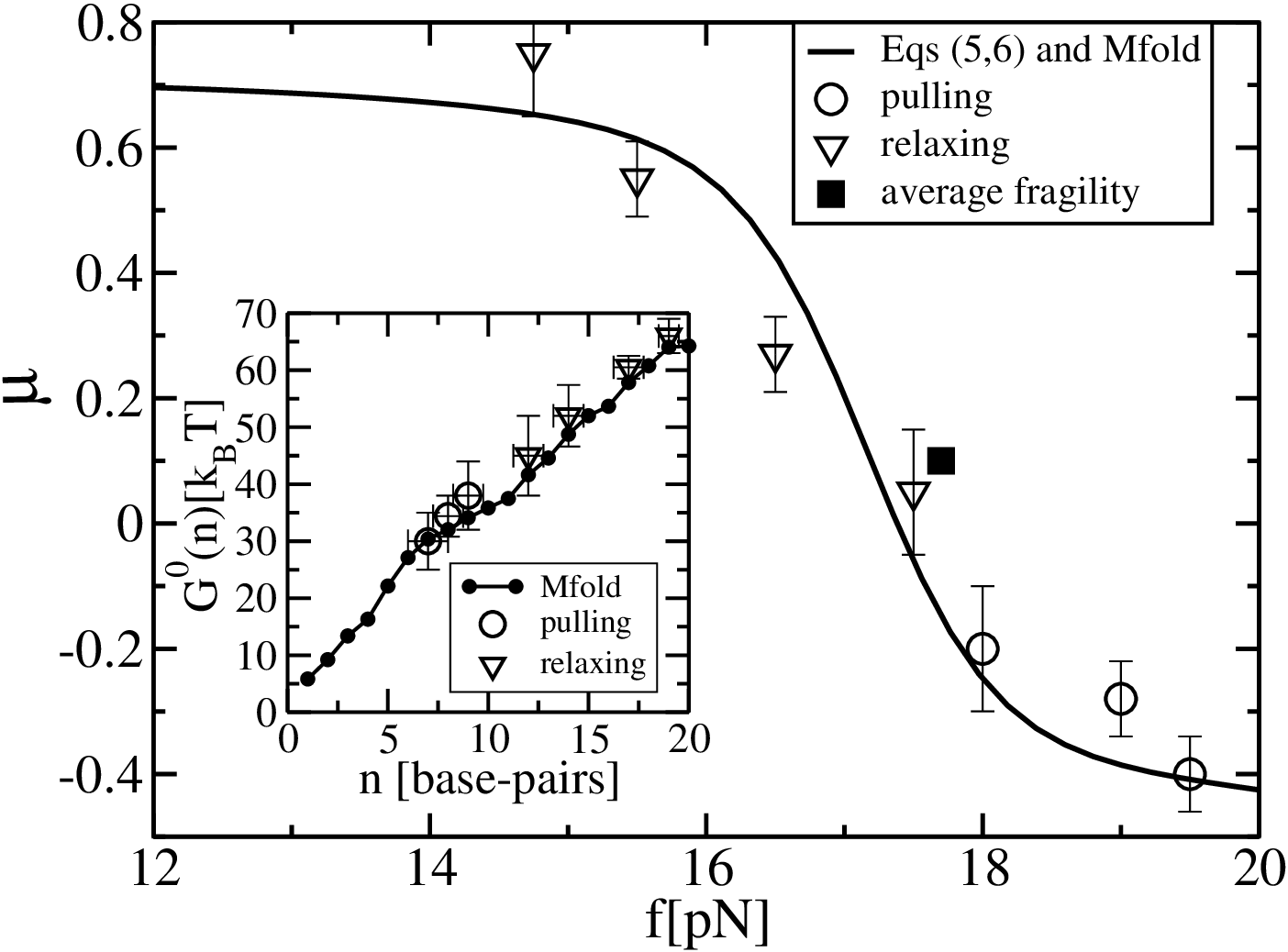}
\caption{\small{Fragility of the RNA hairpin as a function of the
applied force. The continuous line is the prediction by Kramers theory
\eqq{xef}{mu} whereas symbols correspond to the values obtained from
pulling and relaxing data. We also show the average value
of the fragility at $f_c$ (filled square). Inset:
Reconstruction of the zero-force free energy landscape $G^{0}(n)$ compared to the 
Mfold prediction \cite{Mfold}. Error bars indicate 
variability in the estimates obtained for different pulling speeds.}}
\label{f5}
\end{center}
\end{figure}

A kinetic barrier is characterized by its location $n^{*}(f)$ and its height
$G_{n^{*}}(f)$. As shown in the inset of Fig.\ref{f3}, the free energy
landscape at force $\tilde{f}$ shows that there are two barriers
corresponding to transition states located at $n_{1}^{*}\approx 15-19$ and
$n_{2}^{*}\approx 6-7$.  At low forces,
$f<\tilde{f}$ (regime I), the highest barrier is located at $n_{1}^{*}$ and
corresponds to the entropy cost associated to the opening of the four
bases loop. Whereas for large forces, $f>\tilde{f}$ (regime II), the
kinetics is governed by the barrier located at $n_{2}^{*}$ at the interface
between the GC and AU rich regions of the hairpin. In our experiments we
observe the two different regimes, I and II, because the crossover
force, $\tilde{f}=17.3\rm pN$, is within the experimentally
accessible range of rupture forces, $[f_{c}\pm 3\rm pN]$ (inset of
Fig.\ref{f0}), where $f_{c}=17.7\rm pN$ is the critical force 
verifying $G_{N}(f_{c})=G_{0}(f_{c})$ in \eq{gn}. In order to 
investigate the unfolding/refolding kinetics 
over a broader range of forces than those accessible in force-ramp 
experiments, force-jump
experiments \cite{pan} could be very helpful.           
From $B_{\rm eff}^{\rm exp}$ and $B_{\rm eff}^{\rm KT}$ we obtain the
fragility $\mu(f)$ by using \eqq{xef}{mu}.  In Fig.\ref{f5} we show the
agreement between the fragility obtained from the experimentally
measured barrier \eqq{befe1}{befe2} and Kramers theory \eq{bef}.
Finally, our method can be used to experimentally reconstruct the free
energy landscape of the molecule from the sole knowledge of the 
breakage force distribution at different loading rates. We first
determine the location of the force dependent transition state
$n^{*}(f)$ from the measured value of $x^{\rm{F(UF)}}_{\rm eff}(f)$
\eq{xef}.  Using the saddle point approximation we then identify the
effective barrier with the largest contribution to the sum appearing
in the r.h.s of \eq{bef}, $B_{\rm eff}^{\rm exp}(f)\approx
{\rm{Max}}_{n} G_{n}(f) =G_{n^{*}}(f)$, where we have taken
$h(n^{*})\approx 1$ \cite{foot4}.  Finally we apply \eq{gn} and
extrapolate the free energy $G_{n^{*}}(f)$ to zero force to obtain
$G_{n^{*}}(f=0)$. In the inset of Fig.\ref{f5} we show the
experimentally reconstructed free energy landscape compared with the
results obtained from Mfold \cite{Mfold}.

Changes in the position of the transition state \eq{xef} along
the reaction coordinate axis with the force can correspond to two
different situations: (i) The location of the transition state $n^{*}$
does not change with force but the extension $x_{n}(f)$ does for all
configurations as predicted in \cite{thirum1};  (ii) The free energy landscape of
the hairpin shows multiple barriers leading to different transition states
depending on the value of the force. This is the case considered in the present study. 
Interestingly the value of $\mu$ varies with force in
the situation (ii) but not in (i).  Therefore $\mu$ is a good
parameter to identify structural changes in the transition state.

A useful analysis of
experimental data for molecular rupture is the two-states model
\cite{mun,ritort} where the position of the kinetic barrier along the
reaction coordinate is fixed: $x^{\rm{F}},x^{\rm{UF}}$ and the fragility,
$\tilde{\mu}= (x^{\rm{F}}-x^{\rm{UF}})/(x^{\rm{F}}+x^{\rm{UF}})$, do not
depend on the force.  In this approximation, $\ln [r(\ln
(1/P_{\rm{u(r)}}))]$ is a straight line as a function of the applied
force \cite{evans}. A plot of the experimental data for $\ln
[r(\ln (1/P_{\rm{u(r)}}))]$ versus force displays a non-zero curvature
(data not shown) indicating a force dependent fragility. 
Moreover, in two-states systems
the dependence of the mean value of the unfolding and refolding forces on the rate $r$ can be
estimated in the experimental regime where 
$\frac{k_{\rm{u(r)}}(f')k_{B}T}{rx^{\rm{F(UF)}}}<<1$ \cite{hummerszabo03}:
\begin{eqnarray}
\langle
f\rangle_{\rm{u}}\propto\frac{k_{B}T}{x^{\rm{F}}}[\ln(r)],~\langle f\rangle_{\rm{r}}\propto\frac{-k_{B}T}{x^{\rm{UF}}}[\ln(r)].\label{f*}
\end{eqnarray}
%
%
By fitting the experimental results to \eq{f*} 
we can estimate values for $x^{\rm{F}},x^{\rm{UF}}$ 
which give $\tilde{\mu}\approx 0$.
This corresponds to a barrier located in the middle between 
the F and UF states, in
disagreement with the free energy landscape shown in Fig.\ref{f3}. Yet,
this value coincides with the average fragility measured over the range
of forces $f\in[15-19\rm pN]$ (Fig.\ref{f5}) suggesting that fragility
estimates obtained by fitting the two-states model to the experimental
data correspond to averages of force dependent fragilities over the
range of breakage forces explored in the experiments.

We have applied Kramers theory  to
investigate the kinetics of unfolding and refolding of an RNA hairpin
under mechanical force.  The analysis of the experimental data for the
unfolding and refolding force distributions allows us to determine the
location of the force-dependent kinetic barrier, the attempt frequency
$k_{0}$ of the hairpin and the free energy landscape of the molecule. 
The method should be applicable to hairpins with
multiple barriers.
The theory presented here may fail to describe the 
unfolding/refolding of the hairpin at low forces and/or high temperatures, where 
breathing configurations are relevant \cite{thirum1} 
and the free-energy landscape becomes multidimensional. 
The presence of force-induced structural changes in molecular transition states is a
general feature of biomolecules typically showing a rugged free
energy landscape. Proper consideration of the force-dependence of the
fragility is crucial to correctly interpret the results from pulling
experiments and to relate force unfolding measurements with thermal
denaturation experiments.

{\bf Acknowledgments.} We are grateful to Bustamante and Tinoco labs
for kindly providing the facilities where experiments were carried out
and S.B.Smith for technical assistance in the tweezers instrument. M.M
acknowledges a grant from the University of Barcelona. D.C. is
supported by NIH grant GM10840 and F.R by the Ministerio de
Educacion y Ciencia in Spain (FIS2004-03545) and Distincio de la
Generalitat de Catalunya.



\begin{thebibliography}{99}

\bibitem{Review} C. Bustamante, Y. R. Chemla, N. R. Forde and D. Izhaky, Ann. Rev. Biochem. {\bf 73} , 705 (2004).


\bibitem{Mariano} M. Carrion-Vazquez et al., Proc. Nat. Acad. Sci. {\bf 96}, 3694 (1999).


\bibitem{bockelmann97} U. Bockelmann, B. Essevaz-Roulet and F. Heslot,
  Phys. Rev. Lett. {\bf 79}, 4489 (1997).

\bibitem{Liphardt01} J. Liphardt, B. Onoa, S. B. Smith, I. Tinoco Jr, and C. Bustamante, Science {\bf 292}, 733 (2001).

\bibitem{fritz98} J. Fritz, A. G. Katopodis, F. Kolbinger and
  D. Anselmetti, Proc. Nat. Acad. Sci. (USA) {\bf 95}, 12283 (1998). 

\bibitem{Agrait01} G. Rubio-Bollinger, S. R. Bahn, N. Agra\"{\i}t,
  K. W. Jacobsen and S. Vieira, Phys. Rev. Lett. {\bf 87}, 026101 (2001). 

\bibitem{EvansWilliams01} E. Evans and P. Williams in Physics of
Biomolecules and Cells, Les Houches, Eds. H. Flyvberg and F. Julicher,
Springer-Verlag (2002).


\bibitem{Smith03} S. B. Smith, Y. Cui and C. Bustamante,
  Methods. Enzymol. {\bf 361}, 134 (2003).


\bibitem{foot1} Pulling experiments were performed at
$25\,^{\circ}\mathrm{C}$ in 100mM Tris HCl, 8.1pH, 1mM EDTA with a
siRNA hairpin that targets the mRNA of the CD4 receptor of the Human
Immunodeficiency Virus.
For details see D. Collin et al., Nature {\bf 437}, 231
(2005).


\bibitem{Manosas05} Experimentally the micropipette is moved at a
constant pulling speed. However, above 10pN (where the molecule
typically unfolds and refolds) the loading rate is approximately
constant and equal to the stiffness of the trap times the pulling
speed. For details see M. Manosas and F. Ritort, Biophys. J. {\bf 88},
3224 (2005).



\bibitem{cocco03} S. Cocco, R. Monasson, and J.F Marko, Eur. Phys. J. E {\bf 10}, 153 (2003).

\bibitem{foot2} The mechanical response of the ssRNA, $F_{\rm ssRNA}(x)$
and $x_{n}(f)$, is described by the worm-like-chain model
\cite{Bustamante94} with a persistence and contour lengths equal to $P_{\rm
ssRNA} =1\rm nm$ and $L_{\rm ssRNA}=0.59$nm per base. 

\bibitem{Nelson} D. K. Lubensky and D. R. Nelson, Phys. Rev. E {\bf
65}, 031917 (2002) .

\bibitem{Bustamante94} C. Bustamante, J. F. Marko, E. G. Siggia,
S. B. Smith, Science {\bf 265}, 1599 (1994).

\bibitem{Mfold} We use the Visual OMP from DNA software to predict
  free energy values (at $25\,^{\circ}\mathrm{C}$, in 0.1M NaCl).  


\bibitem{Zwanzig} R. Zwanzig, {\it Nonequilibrium Statistical Physics}, 1st Ed. 
(Oxford University Press, 2001), Chapter 4.


\bibitem{leffler} The fragility $\mu$ is directly related to the
parameter $\alpha$ ($\mu=2\alpha-1$) introduced in J. E. Leffler,
Science {\bf 117}, 340 (1953) to characterize the resemblance of the
transition state to the reactant and product of a chemical reaction.

\bibitem{evans} E. Evans and K. Ritchie, Biophys. J. {\bf 72}, 1541 (1997). 


\bibitem{thirum} D. Thirumalai and C. Hyeon, Biochemistry {\bf 44},
 4957 (2005).

\bibitem{foot5} $k_0$ is not the real attempt frequency of the RNA
molecule but has contributions from the setup (handles and bead). Yet
these can be shown that do not to change the order of magnitude of its
value (M. Manosas et al., unpublished).

\bibitem{pan} P.T.X. Li et al.,  Biophys. J. {\bf 90}, 250 (2006).


\bibitem{foot4}
In the experimentally accessible range of rupture forces (around $f_{c}$),
the condition $G_{n}(f)>0$ for $n=1,..,n^{*}$ is verified. Therefore
only the $n'=0$ term of the sum appreciably contributes to $h(n^*)$ in \eq{bef}

\bibitem{thirum1} C. Hyeon and D. Thirumalai, Proc. Natl. Acad. 
Sci. USA {\bf 102}, 16789 (2005); Biophys. J, doi:10.159/biophysj.105.078030 
(2006).


\bibitem{mun} V. Mu\~noz, P.A. Thompson, J. Hofrichter, and W. A. Eaton, Nature {\bf 390}, 196 (1997).


\bibitem{ritort} F. Ritort, C. Bustamante and I.N.Tinoco Jr, Proc. Natl. Acad. Sci. USA {\bf 99}, 13544 (2002).

\bibitem{hummerszabo03} G. Hummer and A. Szabo, Biophys. J. {\bf 85}, 5 (2003).

\end{thebibliography}
\end{document}